\documentclass[a4paper,11pt]{article}

\usepackage{pos}

\usepackage[nodisplayskipstretch]{setspace}
\AtBeginDocument{%
  \addtolength\abovedisplayskip{-0.5\baselineskip}%
  \addtolength\belowdisplayskip{-0.5\baselineskip}%
  \setlength{\belowcaptionskip}{-10pt}
}
\usepackage{titlesec}
\titlespacing\section{0pt}{11pt plus 4pt minus 2pt}{4pt plus 2pt minus 2pt}
\titlespacing\subsection{0pt}{11pt plus 4pt minus 2pt}{4pt plus 2pt minus 2pt}
\titlespacing\subsubsection{0pt}{11pt plus 4pt minus 2pt}{4pt plus 2pt minus 2pt}

\title{New Public Neutrino Alerts for Clusters of IceCube Events }

\ShortTitle{IceCube Public GFU-Cluster Alerts}

\author{The IceCube Collaboration \\{\normalsize \normalfont(a complete list of authors can be found at the end of the proceedings)}\\}

\emailAdd{sarah.mancina@icecube.wisc.edu}
\emailAdd{sergio.cuencalobo@studenti.unipd.it}
\emailAdd{elisa.bernardini@unipd.it}

\abstract{
The IceCube Neutrino Observatory searches for the origins of astrophysical neutrinos using various techniques to overcome the significant backgrounds produced by cosmic-ray air showers. 
One such technique involves combining the neutrino data with other cosmic messengers to identify spatial and temporal correlations. 
IceCube contributes to multi-messenger astrophysics (MMA) by providing alerts for interesting events observed in the detector. 
The Gamma-ray Follow-Up (GFU) cluster alert system is one stream that identifies potential neutrino flares in realtime, producing around 20 alerts per year. 
GFU-cluster alerts have been privately shared with Imaging Air Cherenkov Telescopes (IACTs) through memoranda of understanding since IceCube’s predecessor, AMANDA. 
To preserve blindness to the full behavior of our data, the current system mutes updates from sources following the initial GFU-cluster alert sent, preventing further updates until the activity drops below the alert threshold. 
With growing knowledge of the potential environments that produce astrophysical neutrinos and to foster open collaboration, the GFU-cluster alerts will shift to be publicly shared. 
Additionally, the new alert platform will provide all above-threshold information such that the source behavior after the initial alert is not obscured. 
The above threshold data will be distributed through an interactive website that will update the community on the status of active GFU-cluster alerts. 
This presentation will introduce the new GFU-cluster platform and the accompanying website, soon to be accessible to the MMA community.

\vspace{4mm}

{\bfseries Corresponding authors:}
Sarah Mancina$^{1*}$, 
Sergio Cuenca$^{2}$, 
Elisa Bernardini$^{2}$

{$^{1}$ \itshape INFN-Padova}\\
{$^{2}$ \itshape Università degli Studi di Padova}\\
$^*$ Presenter
}

\ConferenceLogo{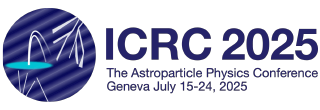}

\FullConference{39th International Cosmic Ray Conference (ICRC2025)\\
 15–24 July 2025\\
Geneva, Switzerland\\}

\begin{document}

\maketitle

\section{Introduction}\label{Intro}

Collaborative efforts between experiments in multi-messenger astrophysics (MMA) are particularly important when studying time-dependent astrophysical phenomena.
Neutrinos can play a unique role in MMA by helping identify the cosmic-ray acceleration processes occurring in extreme astrophysical environments.
The IceCube Neutrino Observatory continuously monitors the sky by detecting signals of Cherenkov radiation from relativistic particles in a cubic kilometer of deep Antarctic ice instrumented with photomultiplier tubes~\citep{IceCube:2016DetectorPaper}.
Its more than 99\% detector uptime and all-sky field of view makes IceCube ideal for sending alerts to the MMA community for follow-up by other instruments.
In 2017, coordination between IceCube and electromagnetic instruments was exemplified when a high-energy neutrino, IC170922A, was found to coincide with a flare from the blazar TXS 0506+056 observed across multiple wavelengths~\citep{IceCube:2018TXSMMA}.

The IceCube realtime platform produces a variety of neutrino alerts, most notably the Gold and Bronze alerts--single high-energy events more likely to be of astrophysical origin~\citep{IceCube:2023IceCat}.
However, alerts based on multiple events have been sent since 2007 by IceCube's predecessor AMANDA-II~\citep{Ackermann:2007cz, IceCube:2016ICRealtimeAlertSys}.
These ``cluster'' alerts offer some advantages over the single-event alerts.
First, the cluster alerts specifically target time-variable activity, providing an estimated duration of the candidate neutrino flare instead of a single timestamp.
Second, by combining the directional information from multiple events, cluster alerts can provide a better source localization.
Finally, the single-event alerts must have high energies (typically $\gtrsim$ 100 TeV) to suppress the atmospheric muon and neutrino background~\citep{IceCube:2023IceCat}.
Instead, the cluster alerts rely on the spatial and timing coincidence of incoming events to distinguish them from the atmospheric backgrounds, which are relatively uniform in time and right ascension.
Therefore, these alerts can be sensitive to sources with softer power-law spectral indices or exponential energy cutoffs.

We focus here on recent improvements to the Gamma-ray Follow-Up (GFU) cluster alerts, which search for potential neutrino flares of up to 180 days.
Originally designed to provide targets of opportunity (ToO) to imaging air Cherenkov telescopes (IACTs), the GFU-cluster alerts use an unbinned maximum likelihood statistical approach to assign a test statistic (TS) to candidate clusters.
The alert stream operates in two modes: a source-list mode that monitors known astrophysical objects and an all-sky mode that scans the regions around all incoming events.
In 2019, the GFU-cluster alerts were upgraded by improving the event selection, adjusting the likelihood technique, adding new sources to the monitoring, and developing the all-sky mode~\citep{Kintscher:2020Thesis}.
In this work, we will focus on a new effort to upgrade the GFU-cluster alerts, focusing on the new dissemination tools for publicly sharing the alerts with the broader MMA community.

The motivation for updating the GFU-cluster alerts and making them public is driven by several key factors.
In general, there has been a shift towards more public sharing of data in the field of MMA because it facilitates cross-collaboration data analysis.
Public alerts would also allow the GFU-cluster alerts to be followed up by a broader range of electromagnetic instruments.
This is particularly motivated by the multi-messenger data of NGC 1068 and Seyfert galaxies, which suggest that at least some neutrino-bright acceleration sites may be opaque to the GeV-TeV gamma rays detected by IACTs~\citep{Kheirandish:2021Seyfert, IceCube:2022NGC1068NT, IceCube:2024NTSeyferts}.
Additionally, the new alert protocol will provide updates on ongoing activity after the initial alerts.
The new GFU-cluster alerts will be shared via NASA GCN (General Coordinates Network)\footnote{\url{https://gcn.nasa.gov/}}~\citep{GCN1995} Notices and a public website hosted by IceCube with information about active and archival alerts.

In section~\ref{GFUAlgo} we will explain the GFU-cluster alert algorithm and changes to the alert sending scheme. 
Then, in section~\ref{AlertDetails} we will discuss what information we propose to share publicly via NASA GCN and the GFU-cluster alerts website for the source-list mode. 
We conclude in section~\ref{Conclusion} with a summary and outlook, including plans for the all-sky mode.

\section{GFU-Cluster Alerts Algorithm}\label{GFUAlgo}

\subsection{Likelihood Approach}
The GFU-cluster alerts use events selected by the GFU event selection, which runs as a filter at the South Pole.
The GFU filter selects for high-quality, track-like events produced by muons which leave long, straight paths of deposited charge in the IceCube detector, resulting in a direction resolution on the scale of 1°~\citep{Kintscher:2020Thesis}.
Events that pass the event selection are promptly transmitted to the northern computing center with the events from other realtime filters.
In the north, each incoming event is assigned an event weight by the source-list and all-sky modes to determine if the full GFU algorithm should be run on the event.

\begin{figure}[tb!]
\centering
\includegraphics[width=0.85\linewidth]{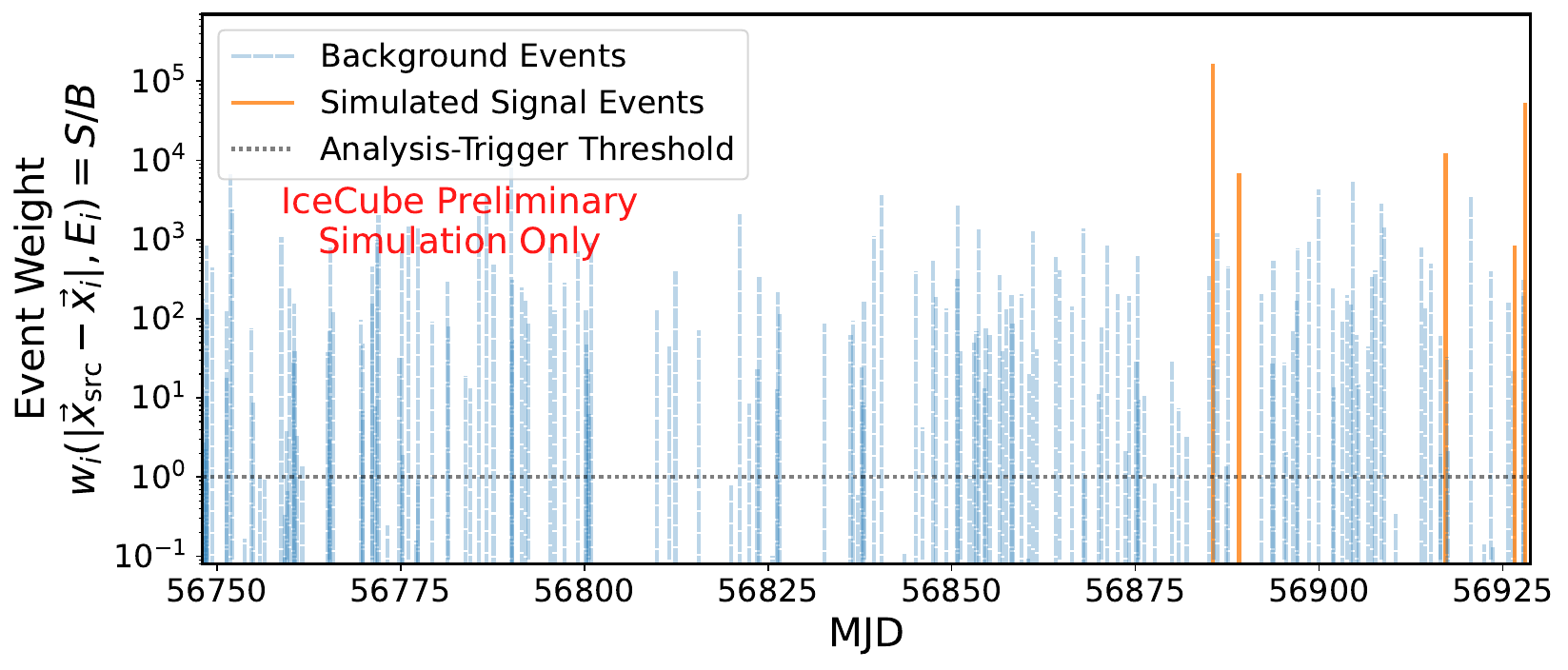}
\caption{Example of the incoming GFU event weights over 180 days (units in Modified Julian Days) around a source at declination -5.69°. The blue lines come from a background simulation derived from real data, with event times randomly shuffled. In orange are 5 signal events that were injected to mimic a flare from the source with a 50-day duration and a spectral index of -2.0. This 180-day period represents the data that the GFU algorithm uses to evaluate the best cluster generated by the final injected event. In the GFU likelihood maximization, $t_1$ is fixed to the time of the latest analysis-triggering event, but fits the time window by allowing the start time, $t_0$ to float between the times of the previous above-threshold events.}\label{fig:eventweight}
\end{figure}
The event weight, $w_i$, is defined as the ratio of the signal to background probability distribution functions (PDF):
\begin{equation}\label{eqn:eventweight}
    w_i = \mathcal{S}(\vec{x}_{\mathrm{src}}, \vec{x}_i, \sigma_i, E_i, \gamma=-3) / \mathcal{B}(\theta_i, \phi_i, E_i).
\end{equation}
The signal PDF, $\mathcal{S}$, is defined as the product of a spatial PDF and an energy PDF.
The spatial term is a Raleigh distribution that is a function of the angular distance between the source location and the event's reconstructed direction, $|\vec{x}_{\mathrm{src}}-\vec{x}_i|$, and the reconstructed angular error, $\sigma_i$.
In the source-list mode, the source locations are pre-defined.
In the all-sky mode, a grid of points within 2° of the incoming event is tested.
The energy term is built from signal simulations and assumes a power-law spectrum ($\propto E^{\gamma}$) assigning a likelihood based on an event's zenith, $\theta_i$, and energy, $E_i$.
In the initial event weight evaluation, the spectral index in the signal PDF is fixed to $\gamma = -3.0$, emphasizing the angular distance to the source over the event's energy.
The spatial and energy PDFs for the background PDF, $\mathcal{B}$, are generated from the data rates under the approximation that most GFU data is produced by atmospheric background and depend upon the event's detector coordinates, zenith ($\theta_i$) and azimuth ($\phi_i$), and the event's energy ($E_i$).
If the event is found to have an event weight greater than one, the full GFU-cluster analysis is run, and we refer to these events as analysis-triggering events.
An example of event weights as a function of Modified Julian Day over 180 days for a simulated source is given in fig.~\ref{fig:eventweight}.

\begin{figure}[tb!]
    \centering
    \includegraphics[width=0.85\linewidth]{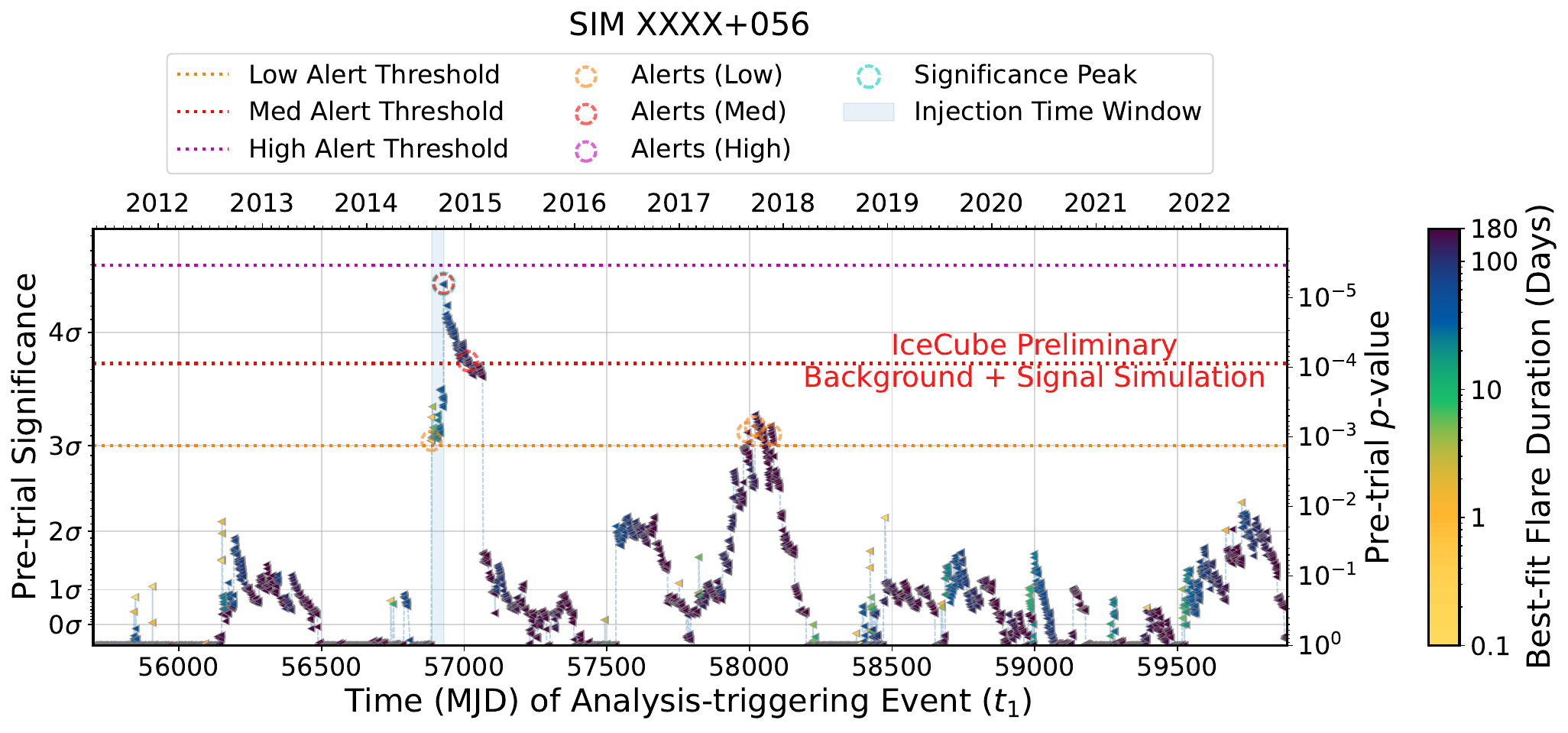}
    \includegraphics[width=0.85\linewidth]{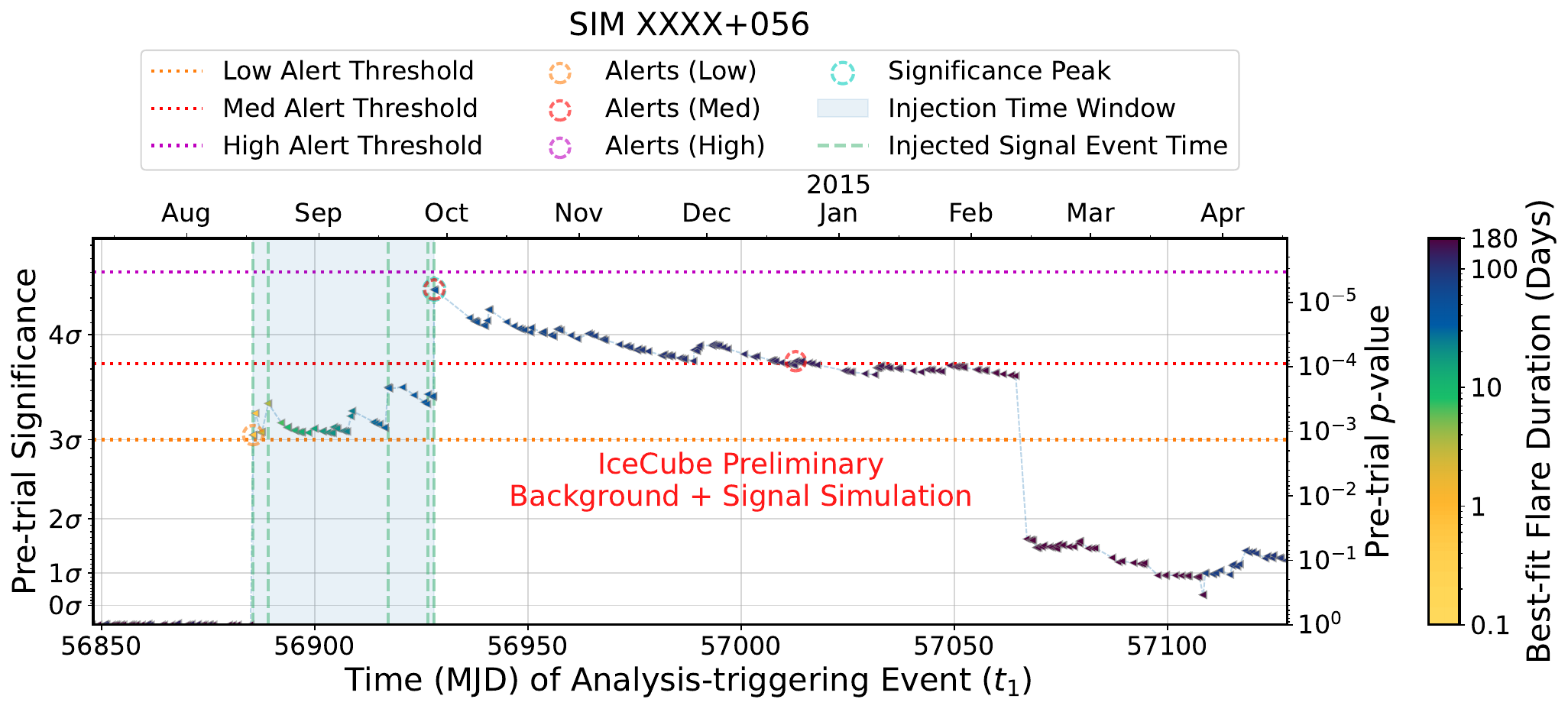}
    \caption{Example significance curve over 11.5 years of data for a simulated source at $\delta$ = 5.69°. The left triangles are the pre-trial significances for clusters from incoming analysis-triggering events plotted at their $t_1$. The color represents the duration of the fitted time window in days ($t_1 - t_0$). Clusters that would trigger alerts by crossing the significance thresholds are highlighted by the dashed circles with colors matching the lines for their respective thresholds. (Top) A source with 5 signal events from a simulated flare with a duration of 50 days and $\gamma = -2.0$ added to simulated background data. The injection time window is highlighted by the light blue area. (Bottom) The significance curve from the top plot is zoomed in around the time of the injected flare. The dashed sea green lines represent the times of the injected signal events (the same events as in fig.~\ref{fig:eventweight}). Note that the significance curve slowly decays over the 180 days following the signal injection.}
    \label{fig:sigcurves}
\end{figure}
When an event triggers the full analysis, the GFU algorithm applies a time-dependent unbinned maximum likelihood fit as is standard in other IceCube point-source analyses~\citep{Braun:2008bg}.
The likelihood takes on the following format:
\begin{equation}\label{eqn:llh}
    \mathcal{L}(n_s, \gamma, t_0; \vec{x}_i, E_i, t_1) = \sum^{N \in [t_0, t_1]}_i \left( \frac{n_s}{N} \mathcal{S}_i + (1-\frac{n_s}{N})\mathcal{B}_i \right)
\end{equation}
where $n_s$ is the fitted number of signal events, $\gamma$ is the fitted spectral index (now allowed to float between -1 and -4), $t_0$ is the fitted start time of the time window, $t_1$ is the fixed time of the incoming analysis-triggering event, and $N$ is the total number of GFU events in the time window defined by [$t_0,t_1$] (including events with $w_i < 1$).
The likelihood is maximized over $n_s$, $\gamma$, and $t_0$.
To reduce computation, the likelihood is maximized over $t_0$ by evaluating only at the previous analysis-triggering event times ($w_i$ > 1) within the past 180 days.

Each analysis-triggering event produces a cluster with best-fit parameters determined by the likelihood maximization.
The test statistic (TS) is calculated by taking the log-likelihood ratio between the likelihood evaluated with the best-fit parameters and the likelihood assuming the null hypothesis, $n_s = 0$, and setting $t_0$ to the best-fit of the numerator term.
A pre-trial p-value can then be assigned to the cluster by comparing the TS to the TS distributions from background simulations generated by scrambling the data in time.
As more clusters come in for a source, a ``significance curve'' can be produced, examples of which are shown in fig.~\ref{fig:sigcurves}.
These curves are not traditional light curves where each point represents a flux measurement over a given time window (all though they appear similar), but rather represent the rejection of the null hypothesis for the cluster evaluated at the time of an incoming event.
Adjacent clusters are not independent and use the same data, and the y-axis is the p-value for only the case of one cluster (i.e., the significance or p-value is not trial-corrected for the total number of clusters).
The pre-trial significance is the metric used to determine if a cluster should be sent as an alert.
To determine the alert threshold, we produce simulations of a year of data and calculate the false alert rate (FAR) as a function of the threshold choice (see fig.~\ref{fig:FAR}).
The FAR as a function of pre-trial significance threshold is much larger for the all-sky alerts than for the source-list alerts due to the larger number of trials from evaluating all incoming events.
However, here we focus only on the source-list alerts, as the initial public alerts will focus on the source-list mode of the GFU-cluster alerts.

\subsection{Current Alert Distribution and Muting Scheme}

In the 2019 version of the GFU-cluster alerts, alerts were shared privately with partner IACTs under Memoranda of Understanding.
These alerts were primarily sent via email and provided details of the first cluster that passed the single alert threshold.
For the source-list alerts, this threshold was chosen to be 3$\sigma$ and result in around 10 alerts per year being sent to each IACT (20 per year overall).
The source behavior was then muted by the system and was only un-muted after an incoming cluster resulted in a pre-trial significance below the alert threshold.
The muting behavior was added to reduce the amount of information ``unblinded'' for these source to reduce bias in offline analyses that were looking at similar source catalogs.
The muting system also prevented spamming our partners with all above threshold clusters but obscured potentially interesting updates to the source after the initial alerts.

\subsection{Proposed Changes to Alert Scheme}
\begin{figure}[tb!]
    \centering
    \includegraphics[width=0.65\linewidth]{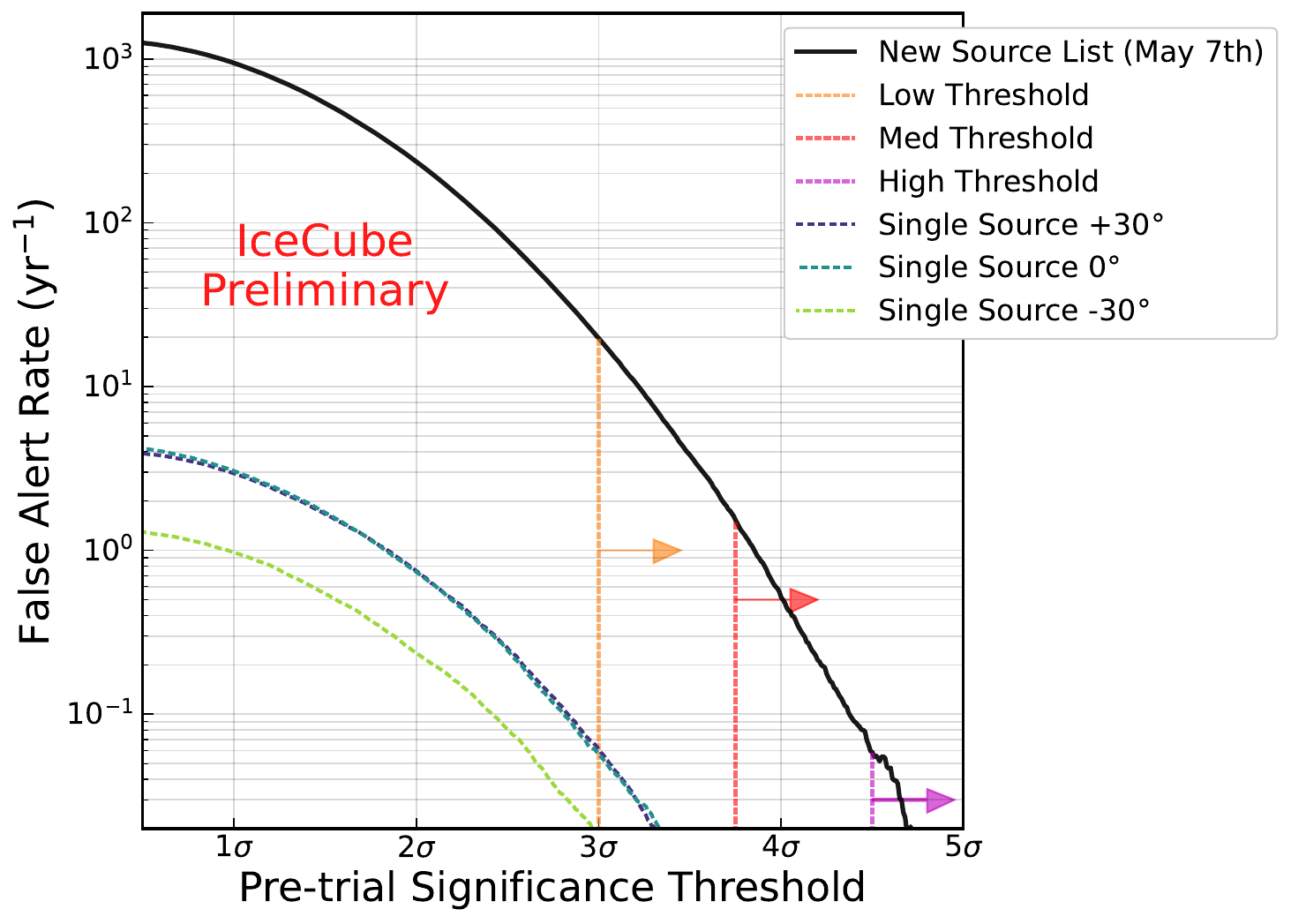}
    \caption{False alarm rate (FAR) from background simulations of 1 year of data as a function of the alert threshold for single sources (dashed lines) and the sum of all sources in the full catalog proposed in~\cite{IceCube:2025CatPreceedings}. The three thresholds proposed for the new alert scheme are highlighted by the dashed vertical lines and arrows.}
    \label{fig:FAR}
\end{figure}
The new GFU-cluster platform will provide updates on the active sources via NASA GCN notice updates and the GFU-cluster alerts website described in section~\ref{AlertDetails}.
The proposed alert scheme will have three alert thresholds: low, medium, and high.
Active sources are sources whose most recent cluster was above the low threshold.
The choice of these thresholds was motivated by the false alarm rate shown in fig.~\ref{fig:FAR}.
The low threshold was chosen to be 3$\sigma$, the same as the previous version of the alert stream, resulting in a FAR of around 20 background alerts per year.
The proposed medium threshold is 3.75$\sigma$, which is expected to produce slightly more than one alert from background per year.
For the high threshold alerts, we selected a threshold of 4.5$\sigma$, which results in around 1 false alert per 20 years.
The different thresholds allow us to update alerts over GCN, notifying the MMA community of the increase in pre-trial significance of the clusters coming from a given source.

\section{Public Alert Content and Distribution Channels}\label{AlertDetails}

\subsection{NASA GCN}
The GFU-cluster alert information sent over GCN will be kept to the basic information for follow-up, and more detailed information will be posted on the website.
We propose to include information about the name and location of the monitored source that produced the alert, the time of the analysis-triggering event ($t_1$) that produced the alert, the best-fit start time of the flare ($t_0$), and the FAR of the cluster converted from the pre-trial significance.
The multi-threshold system can then be used to generate updates to the initial GCN Notice.
In addition, we propose sending GCN Circulars for alerts that surpass the medium and high alert thresholds.

\subsection{GFU-Cluster Website for Updates in Realtime}
\begin{figure}
    \centering
    \includegraphics[width=0.95\linewidth]{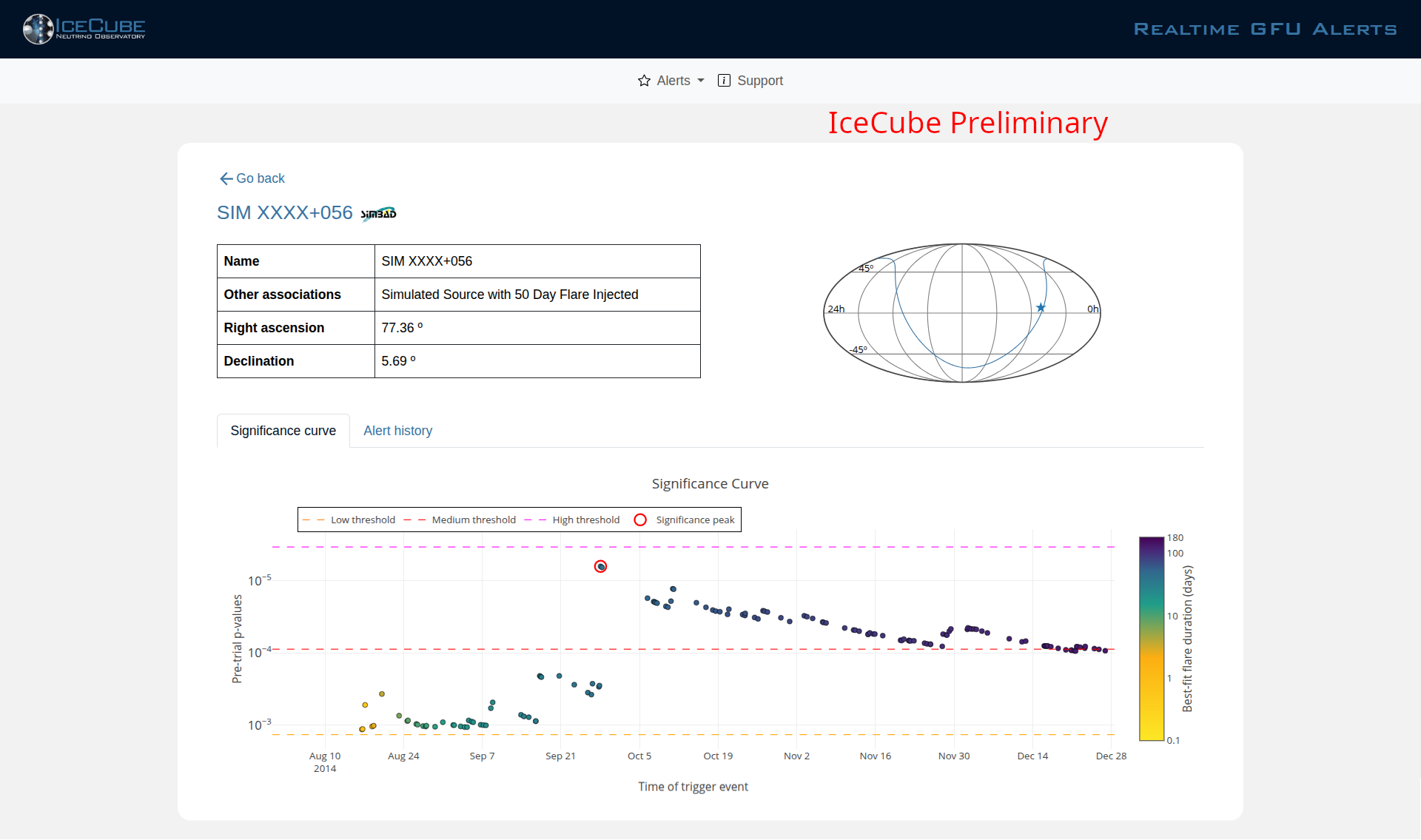}
    \caption{Example source detail page from the GFU-cluster alerts webpage. The source location and details are shared at the top, and the significance curve (note: this is the same simulated source as shown in fig.~\ref{fig:sigcurves}). The significance curve figure is dynamic and can zoom into regions of interest, as is shown here. The source detail page also includes a tab to show a table that stores the source's alert history.}
    \label{fig:webpagesrc}
\end{figure}
The GFU-cluster alerts website will provide more detailed reports on active sources and archival alerts.
This website will be updated as new clusters are evaluated, providing the up-to-date significance curves for the active sources.
The website is built using the Flask framework for building Python-based website applications~\citep{FlaskDoc}.
The main page of the website is the active alerts table, which provides information about any source whose most recent cluster has a pre-trial significance above the low alert threshold.
In addition, an archival alerts page will store all previous alerts from the source.

Clicking on one of the sources in the tables of the active alerts or archival alerts pages takes the user to the source detail page (example shown in fig.~\ref{fig:webpagesrc}).
The source detail page gives details about the source, including the primary name as well as other associated names for the source in the catalog and the location of the source in equatorial coordinates.
Under the source detail, a dynamic plot of the significance curve is presented and will be updated as clusters are evaluated by the GFU-cluster algorithm.
In addition, the source detail page includes the alert history tab, which provides a table with the full history of alerts from the source.
The website will also be documented via the support page.

\section{Conclusion and Outlook}\label{Conclusion}

The GFU-cluster alerts provide the MMA community with neutrino alerts based on clusters of multiple candidate neutrino events.
In this work, we focused on the GFU-cluster source-list alerts, which monitor known astrophysical objects for signs of neutrino flares.
The alerts use a time-dependent unbinned maximum likelihood approach to evaluate the pre-trial significance for clusters of GFU events in realtime.
The pre-trial significance is used as a threshold for sending alerts and is converted into a false alarm rate.
The current version of the GFU alerts has been operating since 2019, and alerts are currently shared privately with IACTs under MoU.

We present a modern update to the GFU-cluster alerts.
We propose sharing the alerts via public channels, over NASA GCN, and on a dedicated website to be hosted by IceCube.
The updated alert stream will remove the old muting scheme, which obscured source behavior after the initial alert, and instead allow for continuous monitoring of the sources.
A multi-threshold scheme will also provide a procedure for sending GCN updates for particularly interesting increases in the significance curve.
In ~\cite{IceCube:2025CatPreceedings}, a detailed discussion of the source list design is presented.
We aim to have the public version of the GFU-cluster source-list alerts available by the end of 2025.
A public version of the all-sky mode is also under production, but its release will follow due to the greater challenge of localizing neutrino sources in that mode.

In the future, neutrino cluster alerts can be used to combine data across multiple neutrino telescopes.
The realtime sphere of neutrino astronomy relies less on years of exposure and instead benefits immediately from any detector's contribution, regardless of its age.
Therefore, joint cluster alerts provide a unique opportunity for fruitful collaboration between IceCube and the many neutrino telescopes under development.

\bibliographystyle{ICRC}
\setlength{\bibsep}{4pt}
\bibliography{references}

%

\clearpage

\section*{Full Author List: IceCube Collaboration}

\scriptsize
\noindent
R. Abbasi$^{16}$,
M. Ackermann$^{63}$,
J. Adams$^{17}$,
S. K. Agarwalla$^{39,\: {\rm a}}$,
J. A. Aguilar$^{10}$,
M. Ahlers$^{21}$,
J.M. Alameddine$^{22}$,
S. Ali$^{35}$,
N. M. Amin$^{43}$,
K. Andeen$^{41}$
% ...
\\
\\
% ...
$^{10}$ Universit{\'e} Libre de Bruxelles, Science Faculty CP230, B-1050 Brussels, Belgium \\
% ...
$^{16}$ Department of Physics, Loyola University Chicago, Chicago, IL 60660, USA \\
$^{17}$ Dept. of Physics and Astronomy, University of Canterbury, Private Bag 4800, Christchurch, New Zealand \\
% ...
$^{21}$ Niels Bohr Institute, University of Copenhagen, DK-2100 Copenhagen, Denmark \\
$^{22}$ Dept. of Physics, TU Dortmund University, D-44221 Dortmund, Germany \\
% ...
$^{35}$ Dept. of Physics and Astronomy, University of Kansas, Lawrence, KS 66045, USA \\
% ...
$^{39}$ Dept. of Physics and Wisconsin IceCube Particle Astrophysics Center, University of Wisconsin{\textemdash}Madison, Madison, WI 53706, USA \\
% ...
$^{41}$ Department of Physics, Marquette University, Milwaukee, WI 53201, USA \\
% ...
$^{43}$ Bartol Research Institute and Dept. of Physics and Astronomy, University of Delaware, Newark, DE 19716, USA \\
% ...
$^{63}$ Deutsches Elektronen-Synchrotron DESY, Platanenallee 6, D-15738 Zeuthen, Germany \\
$^{\rm a}$ also at Institute of Physics, Sachivalaya Marg, Sainik School Post, Bhubaneswar 751005, India \\
% ...

\subsection*{Acknowledgments}

\noindent
The authors gratefully acknowledge the support from the following agencies and institutions:
USA {\textendash} U.S. National Science Foundation-Office of Polar Programs,
U.S. National Science Foundation-Physics Division,
U.S. National Science Foundation-EPSCoR,
U.S. National Science Foundation-Office of Advanced Cyberinfrastructure,
Wisconsin Alumni Research Foundation,
Center for High Throughput Computing (CHTC) at the University of Wisconsin{\textendash}Madison,
Open Science Grid (OSG),
Partnership to Advance Throughput Computing (PATh),
Advanced Cyberinfrastructure Coordination Ecosystem: Services {\&} Support (ACCESS),
Frontera and Ranch computing project at the Texas Advanced Computing Center,
U.S. Department of Energy-National Energy Research Scientific Computing Center,
Particle astrophysics research computing center at the University of Maryland,
Institute for Cyber-Enabled Research at Michigan State University,
Astroparticle physics computational facility at Marquette University,
NVIDIA Corporation,
and Google Cloud Platform;
Belgium {\textendash} Funds for Scientific Research (FRS-FNRS and FWO),
FWO Odysseus and Big Science programmes,
and Belgian Federal Science Policy Office (Belspo);
Germany {\textendash} Bundesministerium f{\"u}r Forschung, Technologie und Raumfahrt (BMFTR),
Deutsche Forschungsgemeinschaft (DFG),
Helmholtz Alliance for Astroparticle Physics (HAP),
Initiative and Networking Fund of the Helmholtz Association,
Deutsches Elektronen Synchrotron (DESY),
and High Performance Computing cluster of the RWTH Aachen;
Sweden {\textendash} Swedish Research Council,
Swedish Polar Research Secretariat,
Swedish National Infrastructure for Computing (SNIC),
and Knut and Alice Wallenberg Foundation;
European Union {\textendash} EGI Advanced Computing for research;
Australia {\textendash} Australian Research Council;
Canada {\textendash} Natural Sciences and Engineering Research Council of Canada,
Calcul Qu{\'e}bec, Compute Ontario, Canada Foundation for Innovation, WestGrid, and Digital Research Alliance of Canada;
Denmark {\textendash} Villum Fonden, Carlsberg Foundation, and European Commission;
New Zealand {\textendash} Marsden Fund;
Japan {\textendash} Japan Society for Promotion of Science (JSPS)
and Institute for Global Prominent Research (IGPR) of Chiba University;
Korea {\textendash} National Research Foundation of Korea (NRF);
Switzerland {\textendash} Swiss National Science Foundation (SNSF).

\end{document}